\documentclass{article}

\PassOptionsToPackage{numbers,sort&compress}{natbib}
\usepackage[dblblindworkshop, final]{neurips_2026}

\workshoptitle{The 6th Workshop on Mathematical Reasoning and AI}

\usepackage[utf8]{inputenc}
\usepackage[T1]{fontenc}
\usepackage{amsmath,amssymb}
\usepackage{booktabs}
\usepackage{graphicx}
\usepackage{multirow}
\usepackage{microtype}
\usepackage{xcolor}
\usepackage{hyperref}
\usepackage{url}
\usepackage{float}

\hypersetup{
  colorlinks=true,
  citecolor=blue,
  linkcolor=blue,
  urlcolor=blue,
  pdfauthor={Kavienan Jegatheesan, Gayathri Lihinikaduarachchi},
  pdfcreator={},
  pdfproducer={}
}

\title{When Tools Lie: Reliability of Mathematical Agents Under Corrupted Tool Feedback}
\author{%
  Kavienan Jegatheesan \\
  Dept.~of Computer Science and Engineering \\
  University of Moratuwa, Sri Lanka \\
  \texttt{kavienanj@gmail.com} \\
  \And
  Gayathri Lihinikaduarachchi \\
  Dept.~of Computer Science and Engineering \\
  University of Moratuwa, Sri Lanka \\
  \texttt{gayathriharshila@gmail.com} \\
}

\begin{document}
\maketitle

\begin{abstract}
Mathematical problem solving often requires deterministic computational steps that agents delegate to tools and implicitly trust. Yet tools can fail silently, returning plausible but incorrect results. How well can agents detect and correct corrupted tool call outputs? We study this through a controlled corruption framework where a hidden interceptor replaces tool call results with plausible incorrect information on targeted problems. We evaluate agents across 31 problems under four verification designs including no verification (baseline), mandatory same-context reflection, optional fresh-context verification, and optional structural verification. Without verification, corruption causes dramatic accuracy loss, from 100\% down to 72.4\%. Mandatory reflection fully recovers this performance to 100\%. Optional verification improves accuracy only when models actively invoke it. Our results show that checking frequency is strongly associated with robustness differences, while unequal invocation prevents a controlled comparison of verifier quality. A supporting recovery experiment shows that full problem restart succeeds in 100\% of cases after explicit detection. These findings demonstrate that verifier availability and verification policy are separate components of mathematical-agent reliability. Mandatory policies enforce verification while optional policies depend on the model's own choice to invoke it.
\end{abstract}

\section{Introduction}

Mathematical agents increasingly use calculators, symbolic solvers, simplifiers, and substitution tools as external evidence. Reliability work commonly treats the model as uncertain and the tool as trusted grounding. Yet software bugs, stale state, integration faults, or adversarial manipulation can instead produce a well-formed, plausible, but false result. Such semantic failures are harder to recognize than timeouts or malformed responses and can be propagated through otherwise coherent reasoning.

Tool-augmented reasoning such as ReAct studies how external actions support reasoning when tools are useful evidence~\citep{yao2022react}. Reflection and self-correction methods revise model-generated reasoning~\citep{shinn2023reflexion,madaan2023self}, although intrinsic self-correction without grounding can be unreliable~\citep{huang2024large}. External checking methods such as CRITIC and Chain-of-Verification introduce additional verification channels~\citep{gou2024critic,dhuliawala2024chain}, while process supervision evaluates intermediate reasoning~\citep{lightman2024let}. We study the complementary setting in which external mathematical evidence itself is plausible but false.

We ask: \textbf{RQ1}, how susceptible are capable mathematical agents to semantic corruption of tool outputs? \textbf{RQ2}, how do no verification, mandatory reflection, optional fresh-context verification, and optional structural verification compare? \textbf{RQ3}, how are checking behavior and conditional success associated with robustness differences? We contribute a controlled corruption framework, an empirical susceptibility study across two capability tiers, and a behavioral decomposition of verification policy. B4 restart is reported only as a supporting recovery observation.

\section{Benchmark and method}

\paragraph{Benchmark and corruption.}
We curated a benchmark drawing from synthetic problems, MATH-500, and GSM8K items spanning algebra, calculus, number theory, and arithmetic. The reported results use a 31-item evaluation subset from this pool, with the models tested identically across all architectures.

\begin{figure}[t]
  \centering
  \includegraphics[width=0.86\linewidth,trim=16 18 16 18,clip]{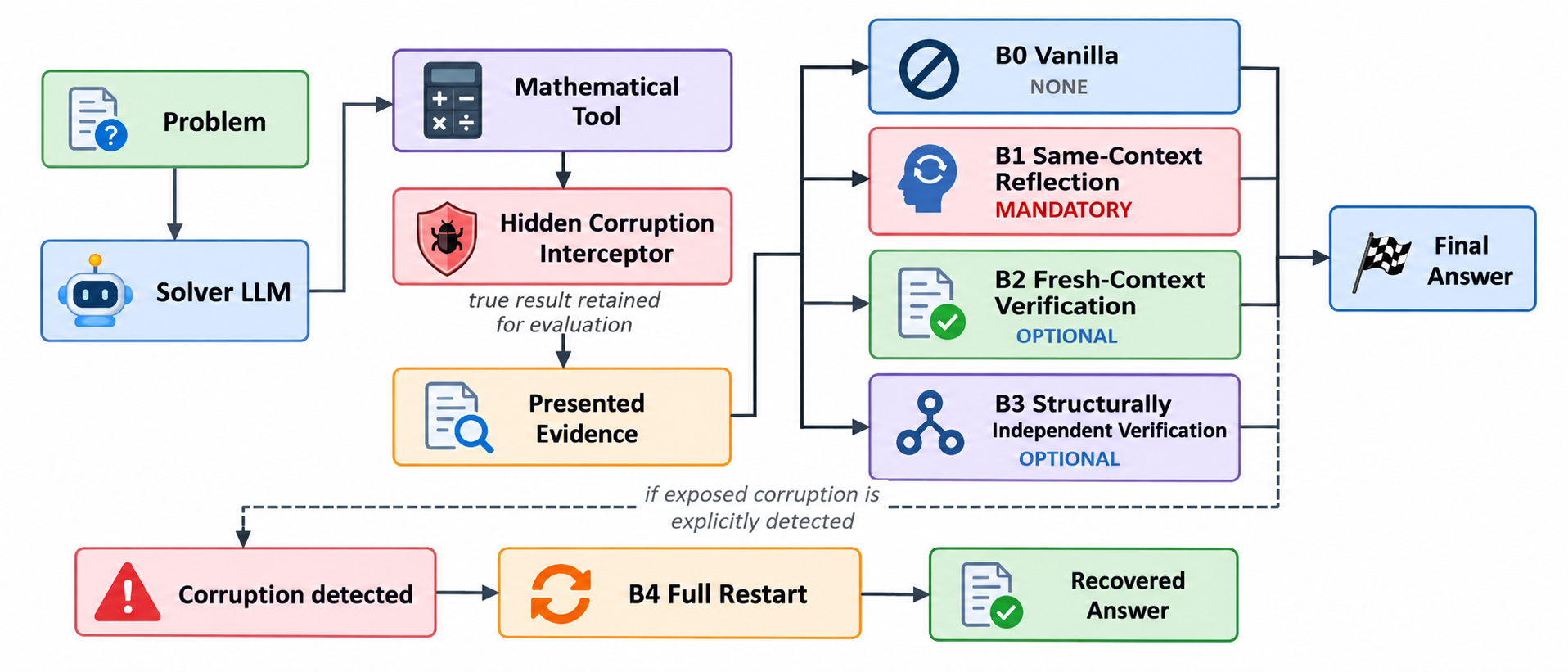}
  \caption{Experimental architecture. A hidden interceptor may replace a genuine tool result with plausible incorrect evidence. B0 has no dedicated verification, B1 mandates same-context reflection, and B2 and B3 provide optional fresh-context and structurally independent verification. B4 restarts only after explicit detection. The key distinction is mandatory checking in B1 versus optional checking in B2/B3.}
  \label{fig:architecture}
\end{figure}

For a corrupted trial, a hidden interceptor replaces the tool's result with incorrect values (Figure~\ref{fig:architecture}). These appear plausible through perturbation, missing solutions, or symbolic changes. Each corruption is independently validated as wrong. Corruption is transient, with at most one tool call corrupted per trial, allowing us to test whether models can detect and correct a single error.

\paragraph{Agent policies and models.}
We test four verification policies illustrated in Figure~\ref{fig:architecture}. B0 (vanilla) uses an ordinary tool-using loop with no verification. B1 (mandatory reflection) intercepts the initial solution submission and unconditionally prompts a same-context reconsideration pass, after which the agent resubmits a possibly revised answer. B2 (optional fresh-context verification) lets the solver ask a fresh model instance to review the problem without the original trace. B3 (optional structural verification) optionally checks using a mathematically different operation, such as testing roots by substitution. Critically, B1--B3 test complete verification policies rather than isolating verifier quality alone. B1 unconditionally prompts a mandatory reconsideration step, whereas B2/B3 test whether the solver chooses to use an available verifier. B4 (full restart) is a supporting post-detection baseline where the solver restarts after explicit corruption detection.

We compare two models from one provider family (Claude Haiku 4.5 and Claude Sonnet 5) to evaluate across capability levels while holding the API environment constant. Both models use identical prompts and fixed parameters across all conditions. The reported results use 496 main-line evaluation trials on the same 31-problem benchmark under all four verification policies.

\paragraph{Evaluation and statistics.}
The primary outcome is final-answer accuracy. We assigned corruption to 248 trials, and corruption reached the agent in 221 of these (89.1\%). The other 27 targeted a tool the agent never invoked and are excluded. Of the 496 reported trials, 488 were graded. The 8 ungraded trials were all Haiku, 7 clean and 1 corrupted B1, and ended without a structured submission. We exclude them rather than score them wrong because their free-text answers all matched the reference. The remaining 220 corrupted trials form Table~\ref{tab:robustness}. Grading normalizes across units, currency, algebraic notation, and LaTeX/Unicode forms to ensure consistent evaluation. Grading is rule-based and deterministic, so the same answer always receives the same grade. To check run-to-run stability, we re-ran B0 twice more on a 15-problem subset for each model, giving three runs per problem and condition. Outcomes agreed across all three runs in 57 of 60 cells (95.0\%), and all three disagreements were in corrupted cells. For statistical analysis, we use problem-level bootstrap 95\% confidence intervals and McNemar tests for the primary B0-versus-B1 comparison per model. Full pairwise comparisons across all policies are reported in Appendix A. A behavioral check is defined as any instance where the model repeats the disputed computation, uses a different mathematical or tool-based cross-check, or invokes a formal verifier. Clean-condition accuracies are computed separately for each verification policy (B0--B3) on graded clean trials. False-distrust rates measure the rate of incorrectly flagging valid tool outputs during verification. These rates are computed over all 31 clean trials per architecture, including those with ungraded final answers, since distrust behavior is observable from transcripts independent of final grading.

\section{Results}

\subsection{RQ1: susceptibility to corrupted evidence}

\begin{figure}[t]
  \centering
  \includegraphics[width=0.86\linewidth,trim=0 0 0 20,clip]{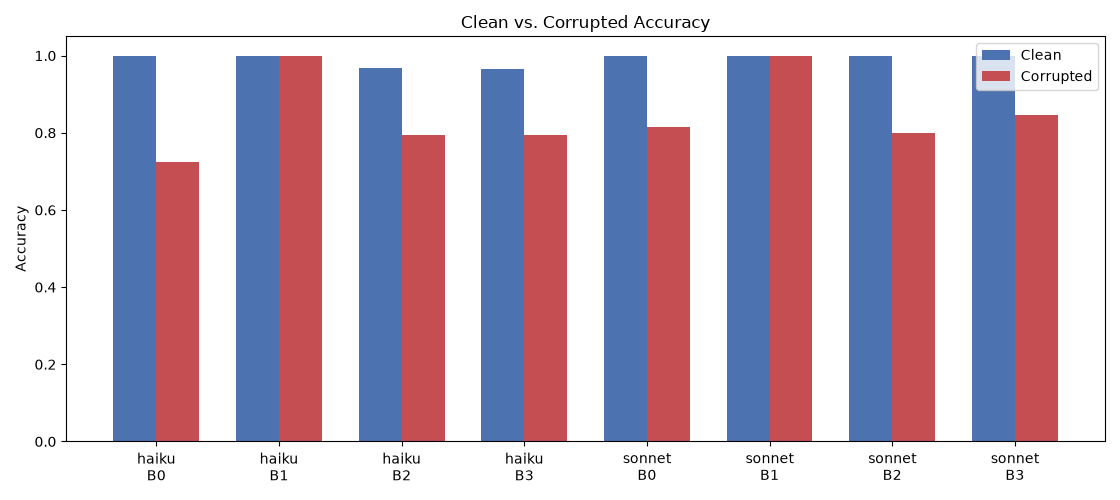}
  \caption{Clean and corrupted final-answer accuracy under B0--B3 for both model tiers. The largest degradation occurs without verification, while mandatory reflection substantially reduces the gap.}
  \label{fig:accuracy}
\end{figure}

The unprotected B0 baseline shows high accuracy on clean items but degrades substantially under plausible corruption (Figure~\ref{fig:accuracy}). Haiku falls from 100\% accuracy on clean items (28/28 graded trials, with 3 ungraded clean trials that ended without a structured submission) to 72.4\% accuracy on corrupted items (21/29 graded trials, 95\% CI: [55.2, 86.2]). This represents a 27.6 percentage-point drop in accuracy. Sonnet shows a similar pattern, falling from 100\% (31/31 clean) to 81.5\% (22/27 corrupted, 95\% CI: [66.7, 96.3]), an 18.5-point drop. Higher clean-task capability does not eliminate vulnerability to false external evidence. Instead, the lower-capability model (Haiku) degrades substantially more under corruption than the higher-capability model (Sonnet).

We further observe that most corruptions occur early in the reasoning process. In 213 of the 221 corrupted trials (96.4\%), corruption occurs on the first tool call. Only 8 trials have corruption on a later tool call. None of these 8 later-position corruptions is explicitly detected, but all 8 come from two Haiku problems and still end with a correct answer, too few to show that later corruption is harder to detect or more harmful. Our results therefore mainly describe corruption at the start of a solution.

\subsection{RQ2: verification-policy comparison}

Mandatory reflection is the strongest end-to-end policy measured. Effect sizes in Table~\ref{tab:robustness} show consistent improvement across both models. Haiku accuracy rises from 72.4\% (B0) to 100\% (B1). Sonnet rises from 81.5\% to 100\%. The Haiku B0-versus-B1 comparison is statistically significant (McNemar $p=0.0078$, $n=28$ paired problems). Sonnet shows a directional trend ($p=0.0625$, $n=26$ paired problems) consistent with Sonnet's already-high B0 baseline leaving less room for a detection effect. Appendix A Table~\ref{tab:pairwise_mcnemar} reports paired-subset accuracies and all pairwise comparisons. These results compare policies with unequal invocation, so they do not show that same-context reflection is intrinsically a better verifier than fresh-context or structural verification. Mandatory reflection introduces a small false-distrust cost (3.2\% for both models, see Appendix A Table~\ref{tab:clean_all_archs}), where correct tool outputs are occasionally flagged as wrong during verification. However, this cost does not reduce final accuracy.

\begin{table}[H]
  \caption{Final-answer accuracy (percentage points) on corrupted trials. Parentheses give correct/total counts, brackets are bootstrap 95\% CIs (Clopper-Pearson exact binomial for 100\% cells to avoid degeneracy). B1 is mandatory, B2/B3 are optional.}
  \label{tab:robustness}
  \centering
  \setlength{\tabcolsep}{3.5pt}
  \begin{tabular}{@{}lcccc@{}}
    \toprule
    Model & B0 none & B1 reflect & B2 fresh & B3 structural \\
    \midrule
    Haiku & 72.4 (21/29) & \textbf{100.0 (28/28)} & 79.3 (23/29) & 79.3 (23/29) \\
          & [55.2,86.2] & \textbf{[87.7,100.0]} & [65.5,93.1] & [65.5,93.1] \\
    \addlinespace[2pt]
    Sonnet & 81.5 (22/27) & \textbf{100.0 (27/27)} & 80.0 (20/25) & 84.6 (22/26) \\
           & [66.7,96.3] & \textbf{[87.2,100.0]} & [64.0,96.0] & [69.2,96.2] \\
    \bottomrule
  \end{tabular}
\end{table}

\subsection{RQ3: checking frequency and robustness}

Pooled across models, B1 checks in 90.9\% of relevant trials (50/55) and succeeds in 100\% of checked cases (50/50). B2 checks in 68.5\% (37/54) and succeeds in 100\% (37/37). B3 checks in 74.5\% (41/55) and succeeds in 95.1\% (39/41). No-check success is 100\% for B1 ($n=5$ trials), 35.3\% for B2 (6/17 trials), and 42.9\% for B3 (6/14 trials). Check classification is based on manual review of trial transcripts. Thus
\begin{equation}
\Pr(S)=\Pr(C)\Pr(S\mid C)+\Pr(\neg C)\Pr(S\mid\neg C),
\end{equation}
and the largest observed separation is checking frequency, $\Pr(C)$, rather than conditional success. This descriptive decomposition does not establish checking as a causal driver of robustness.

Of the 55 graded corrupted B1 trials, 39 made a different tool call after the corrupted result, 11 repeated the identical call, and 5 made no further call. Because corruption is transient, the 11 repeats received the genuine value, so part of B1's recovery may reflect retry rather than detection. The optional formal verifier was invoked in only 7 of 54 corrupted B2 trials and 19 of 55 corrupted B3 trials, and the final answer was correct in 7 of 7 and 18 of 19 of these. Explicit detection rates should be interpreted as lower bounds on corrective behavior because some agents fix the answer through silent recomputation without flagging the original error.

\paragraph{Supporting recovery result.}
Full problem restart succeeds in 100\% of cases after explicit detection (85 attempts, 81 graded). However, this is limited to short-horizon tasks where corruptions occur early.

\section{Discussion and limitations}

The results distinguish conditional performance from utilization. B2/B3 perform well when checking occurs but remain closer to B0. Their checks are optional, whereas B1 guarantees reconsideration. Checking frequency is thus a prominent correlate, not a causal explanation. The high success of invoked verifiers in RQ3 suggests that invocation rather than verifier quality limits B2/B3, but these subsets are small and self-selected. Agents may call a verifier mainly when an error is already easy to spot, so its success on those trials may overstate how well it would perform if invoked every time. A forced condition that always invokes the B2 or B3 verifier would compare verifier quality at equal invocation, and we leave it to future work.

Several limitations constrain generality. Small sample sizes (cells below 30 observations due to the 31-problem pilot benchmark) limit statistical power. Ceiling effects on B1 (100\% for both models) prevent determining whether mandatory reflection is optimal. Evaluation uses only one provider family, limiting cross-provider generalization. The problems are also easy enough to recompute mentally, so B1's 100\% recovery may not carry over to problems where the tool is genuinely needed. Transient corruption also makes recovery easier than a persistently faulty tool would. Because checking is unrandomized, the frequency-robustness association is correlational, not causal.

\section{Conclusion}

Tool-augmented mathematical agents are vulnerable to plausible false tool results despite high clean-task accuracy. Verification policy is a major determinant of robustness in our experiments, with optional policies limited in part by under-utilization of checking. Mandatory reflection recovers accuracy to 100\%, while optional verification remains degraded at 80\% because models underutilize checking. Because verifier invocation differs across conditions, our results do not isolate verifier quality itself. These results highlight a critical distinction for deployment. Building better verifiers is necessary but insufficient. Effective policies must guarantee verification invocation or implement triggering mechanisms that reliably activate verification under risk.

\appendix

\section{Clean-condition accuracy and false distrust across all architectures}

Clean-condition accuracy (and false-distrust rates, FDR) are reported for all four solver architectures in Table~\ref{tab:clean_all_archs}. Clean trials use the same 31-problem evaluation set as corrupted trials. FDR measures the false-rejection rate under clean conditions (the fraction of correct tool outputs flagged as wrong during verification, a false positive). Accuracy is computed using graded trials only (excluding the 8 trials that ended without a structured submission), while FDR is computed over all 31 clean trials per architecture, including those with ungraded final answers, since distrust behavior is observable from transcripts independent of final grading. Haiku's B2 and B3 achieve slightly lower clean accuracy (96.7\%, 96.6\%) than B0/B1, while Sonnet maintains 100\% across all architectures. B1's mandatory reflection creates a measurable but small FDR (3.2\% for both models), while optional verification in B2/B3 incurs no additional false distrust (0\% for both models).

Pairwise McNemar comparisons across all corrupted-condition trials are reported in Table~\ref{tab:pairwise_mcnemar}. The primary pre-specified comparison (B0 vs. B1, per model) is reported in the main text. Comparisons beyond this are exploratory and reported here without multiplicity correction. Accuracies in Table~\ref{tab:pairwise_mcnemar} are computed on the intersection (paired subset) of problems where both architectures within a comparison have exposed trials, and may differ slightly from Table~\ref{tab:robustness}'s marginal accuracies, which report each architecture's full exposed-trial set (e.g., 72.4\% for Haiku B0 on 29 trials vs. 71.4\% on the 28-problem paired subset). Haiku shows significant improvement from B0 to B1 ($p=0.0078$) and directional reversal from B1 to B2 (B1 higher, $p=0.0312$). B2 and B3 show no detectable difference ($p=1.0$). Sonnet shows directionally similar patterns (B0 vs B1 $p=0.0625$, B1 vs B2 $p=0.0625$, B2 vs B3 $p=1.0$) but does not reach significance at these sample sizes given Sonnet's already-high B0 baseline.

\begin{table}[H]
  \caption{Clean-condition accuracy and false-distrust rate (FDR) across all architectures. Parentheses give correct/total counts. FDR is the false-rejection rate on correct tool outputs. Accuracy denominators are graded trials only. FDR denominators include all 31 clean trials per architecture. B0 has no verification step, so FDR is N/A.}
  \label{tab:clean_all_archs}
  \centering
  \setlength{\tabcolsep}{4pt}
  \begin{tabular}{@{}lcccccccc@{}}
    \toprule
    & \multicolumn{4}{c}{Clean Accuracy} & \multicolumn{4}{c}{False Distrust Rate} \\
    \cmidrule(lr){2-5} \cmidrule(lr){6-9}
    Model & B0 & B1 & B2 & B3 & B0 & B1 & B2 & B3 \\
    \midrule
    Haiku & \textbf{100.0} & \textbf{100.0} & 96.7 & 96.6 & N/A & 3.2\% & 0.0\% & 0.0\% \\
          & (28/28) & (30/30) & (29/30) & (28/29) & — & (1/31) & (0/31) & (0/31) \\
    \addlinespace[2pt]
    Sonnet & \textbf{100.0} & \textbf{100.0} & \textbf{100.0} & \textbf{100.0} & N/A & 3.2\% & 0.0\% & 0.0\% \\
           & (31/31) & (31/31) & (31/31) & (31/31) & — & (1/31) & (0/31) & (0/31) \\
    \bottomrule
  \end{tabular}
\end{table}

\begin{table}[H]
  \caption{Pairwise McNemar tests on corrupted trials. Columns give the number of paired problems, accuracies in each architecture, accuracy difference (B minus A), and McNemar $p$-value. Significant results ($p \le 0.05$) in bold.}
  \label{tab:pairwise_mcnemar}
  \centering
  \setlength{\tabcolsep}{5pt}
  \begin{tabular}{@{}lcccccr@{}}
    \toprule
    Model & Comparison & $n$ pairs & Acc. A & Acc. B & $\Delta$ Acc. & $p$-value \\
    \midrule
    \multirow{3}{*}{Haiku} & B0 vs B1 & 28 & 71.4\% & 100.0\% & +28.6pp & \textbf{0.0078} \\
    & B1 vs B2 & 28 & 100.0\% & 78.6\% & $-$21.4pp & \textbf{0.0312} \\
    & B2 vs B3 & 29 & 79.3\% & 79.3\% & 0 & 1.0 \\
    \addlinespace[2pt]
    \multirow{3}{*}{Sonnet} & B0 vs B1 & 26 & 80.8\% & 100.0\% & +19.2pp & 0.0625 \\
    & B1 vs B2 & 24 & 100.0\% & 79.2\% & $-$20.8pp & 0.0625 \\
    & B2 vs B3 & 25 & 80.0\% & 84.0\% & +4.0pp & 1.0 \\
    \bottomrule
  \end{tabular}
\end{table}

\section{Prompt Templates and System Instructions}

The prompts control whether verification is mandatory, optional, or absent.

\subsection*{B0 -- Vanilla (No Verification)}
\noindent\fcolorbox{black}{gray!15}{%
  \begin{minipage}{\dimexpr\linewidth-2\fboxsep\relax}
    \small\itshape
    You are a careful mathematical problem solver with access to tools. Use the appropriate tool to compute results rather than doing arithmetic or algebra by hand. When you have your final answer, call submit\_final\_answer -- that is the only way to finish the problem.
  \end{minipage}%
}

\subsection*{B1 -- Mandatory Same-Context Reflection}

\noindent\textbf{B1 System Prompt (Initial)}\\[0.3em]
\noindent\fcolorbox{black}{gray!15}{%
  \begin{minipage}{\dimexpr\linewidth-2\fboxsep\relax}
    \small\itshape
    You are a careful mathematical problem solver with access to tools. Use the appropriate tool to compute results rather than doing arithmetic or algebra by hand. When you have your final answer, call submit\_final\_answer -- that is the only way to finish the problem.
  \end{minipage}%
}

\vspace{0.5em}

\noindent\textbf{B1 Reflection Prompt (After First Submit)}\\[0.3em]
\noindent\fcolorbox{black}{gray!15}{%
  \begin{minipage}{\dimexpr\linewidth-2\fboxsep\relax}
    \small\itshape
    Before finalizing, check whether your reasoning, tool use, and final answer are correct. If anything looks inconsistent, use \texttt{challenge\_evidence} to flag it and recompute as needed. Then call \texttt{submit\_final\_answer} again with your (possibly revised) answer.
  \end{minipage}%
}

\subsection*{B2 -- Optional Fresh-Context Verification}

\noindent\textbf{B2 Solver Prompt}\\[0.3em]
\noindent\fcolorbox{black}{gray!15}{%
  \begin{minipage}{\dimexpr\linewidth-2\fboxsep\relax}
    \small\itshape
    You are a careful mathematical problem solver with access to tools. Use the appropriate tool to compute results rather than doing arithmetic or algebra by hand. You may use verify\_with\_fresh\_reviewer to have an independent reviewer double-check a computed result before relying on it. When you have your final answer, call submit\_final\_answer -- that is the only way to finish the problem.
  \end{minipage}%
}

\vspace{0.5em}

\noindent\textbf{B2 Verifier (Fresh Instance) Prompt}\\[0.3em]
\noindent\fcolorbox{black}{gray!15}{%
  \begin{minipage}{\dimexpr\linewidth-2\fboxsep\relax}
    \small\itshape
    You are an independent mathematical reviewer. You will be shown a problem and a single claimed result from a calculation tool. You do not know how the claim was produced or who produced it, and you must not assume it is correct -- check it directly against the original problem and report your verdict.
  \end{minipage}%
}

\noindent The verifier observes only the original problem and the disputed claim, not the solver's reasoning.

\subsection*{B3 -- Optional Structural Verification}

\noindent\textbf{B3 Solver Prompt}\\[0.3em]
\noindent\fcolorbox{black}{gray!15}{%
  \begin{minipage}{\dimexpr\linewidth-2\fboxsep\relax}
    \small\itshape
    You are a careful mathematical problem solver with access to tools. Use the appropriate tool to compute results rather than doing arithmetic or algebra by hand. You may use verify\_tool\_result to independently double-check a computed result before relying on it. When you have your final answer, call submit\_final\_answer -- that is the only way to finish the problem.
  \end{minipage}%
}

\section{Check Classification Criteria}

A behavioral \emph{check} is recorded when the trial transcript shows explicit verification action. Classification is based on manual review of agent reasoning traces. Three categories are recognized:

\subsection*{Recomputation}
Agent repeats the disputed tool call or computation to independently confirm the result.

\noindent\fcolorbox{black}{gray!15}{%
  \begin{minipage}{\dimexpr\linewidth-2\fboxsep\relax}
    \small\itshape
    Example (Sonnet, B1). Asked to evaluate 2*x**2 - 2 at x = 3, the agent receives the corrupted value 88/5 and flags it with ``2*3**2 - 2 = 2*9-2=16, not 88/5. Tool result appears incorrect.'' It then repeats the identical substitution, receives 16, and submits 16.
  \end{minipage}%
}

\subsection*{Cross-Check}
Agent uses a different mathematical or tool-based method to verify a claimed result without replication.

\noindent\fcolorbox{black}{gray!15}{%
  \begin{minipage}{\dimexpr\linewidth-2\fboxsep\relax}
    \small\itshape
    Example (Haiku, B1). Asked to solve x**2 - 11*x + 24 = 0, the agent receives only the root 8. It writes ``Let me verify this result, as a quadratic equation typically has two solutions.'', factors the expression as (x - 3)(x - 8), substitutes both roots, and reports ``Both solutions check out.'' before submitting 3, 8.
  \end{minipage}%
}

\subsection*{Formal Verification}
Agent invokes a formal verifier, either \texttt{verify\_tool\_result} (B3) or \texttt{verify\_with\_fresh\_reviewer} (B2), to independently validate a claim.

\noindent\fcolorbox{black}{gray!15}{%
  \begin{minipage}{\dimexpr\linewidth-2\fboxsep\relax}
    \small\itshape
    Example (Sonnet, B3). Asked to evaluate 3*x**2 - 9*x - 5 at x = 3, the agent receives the corrupted value -9/2 and calls \texttt{verify\_tool\_result}, which returns FAIL. It then flags the result with ``Manual calculation: 3*3\^{}2 - 9*3 - 5 = 27 - 27 - 5 = -5, not -9/2.'' and submits -5.
  \end{minipage}%
}

\section{Example Corrupted Trial Comparisons}

The two trials below are taken directly from the experiment logs. Within each trial, all four policies received the same corrupted value. Quoted text is verbatim model output with Markdown formatting removed.

\subsection*{Logged Trial 1: Definite Integral (Haiku)}

\noindent\fcolorbox{black}{blue!10}{%
  \begin{minipage}{\dimexpr\linewidth-2\fboxsep\relax}
    \small
    \textbf{Problem:} ``Compute the definite integral of 4*x**2 from 1 to 3.''

    \textbf{Genuine result:} $\int_1^3 4x^2\,dx = \tfrac{4}{3}(27 - 1) = 104/3$.

    \textbf{Corruption:} The integration tool returns 156/5 instead of 104/3. This numeric perturbation was labeled \emph{plausible} by our severity rules.
  \end{minipage}%
}

\vspace{0.5em}
\noindent\textbf{B0 (Vanilla -- No Verification).}

\noindent\fcolorbox{black}{gray!15}{%
  \begin{minipage}{\dimexpr\linewidth-2\fboxsep\relax}
    \small\itshape
    The agent reports ``The definite integral of 4*x**2 from 1 to 3 is 156/5 or 31.2.'' and submits. \textbf{Final answer: 156/5 (incorrect).}
  \end{minipage}%
}

\noindent\textbf{B1 (Mandatory Reflection).}

\noindent\fcolorbox{black}{gray!15}{%
  \begin{minipage}{\dimexpr\linewidth-2\fboxsep\relax}
    \small\itshape
    The agent first submits 156/5. After the reflection prompt, it checks the antiderivative $\tfrac{4}{3}x^3$ by differentiating it, evaluates it at both bounds, and flags the tool result with the reason ``Manual calculation using antiderivative (4/3)*x**3 evaluated from 1 to 3 gives 36 - 4/3 = 104/3, but the tool returned 156/5. These don't match.'' \textbf{Final answer: 104/3 (correct).} The check uses a different method rather than repeating the integration.
  \end{minipage}%
}

\noindent\textbf{B2 (Optional Fresh-Context Verification).}

\noindent\fcolorbox{black}{gray!15}{%
  \begin{minipage}{\dimexpr\linewidth-2\fboxsep\relax}
    \small\itshape
    The agent reports ``The definite integral of 4*x**2 from 1 to 3 is 156/5 or 31.2.'' and submits without calling the fresh reviewer. \textbf{Final answer: 156/5 (incorrect).}
  \end{minipage}%
}

\noindent\textbf{B3 (Optional Structural Verification).}

\noindent\fcolorbox{black}{gray!15}{%
  \begin{minipage}{\dimexpr\linewidth-2\fboxsep\relax}
    \small\itshape
    Without calling the structural verifier, the agent recomputes the integral by hand, writes ``Hmm, I got 104/3, but the tool returned 156/5.'', and rejects the tool result. \textbf{Final answer: 104/3 (correct).} The recovery comes from a spontaneous check, not from the available verifier.
  \end{minipage}%
}

\subsection*{Logged Trial 2: Modular Arithmetic (Sonnet)}

\vspace{0.5em}
\noindent\fcolorbox{black}{blue!10}{%
  \begin{minipage}{\dimexpr\linewidth-2\fboxsep\relax}
    \small
    \textbf{Problem:} ``What is 84 * 86 + 16, modulo 19?''

    \textbf{Genuine result:} $84 \times 86 + 16 = 7240$, and $7240 = 19 \times 381 + 1$, so the answer is 1.

    \textbf{Corruption:} The first modular evaluation returns 0 instead of 1. This numeric perturbation was labeled \emph{obvious} by our severity rules.
  \end{minipage}%
}

\vspace{0.5em}
\noindent\textbf{B0 (Vanilla -- No Verification).}

\noindent\fcolorbox{black}{gray!15}{%
  \begin{minipage}{\dimexpr\linewidth-2\fboxsep\relax}
    \small\itshape
    The agent reports ``The answer is 0.'' and submits. \textbf{Final answer: 0 (incorrect).}
  \end{minipage}%
}

\noindent\textbf{B1 (Mandatory Reflection).}

\noindent\fcolorbox{black}{gray!15}{%
  \begin{minipage}{\dimexpr\linewidth-2\fboxsep\relax}
    \small\itshape
    The agent first submits 0. After the reflection prompt, it flags the tool result with the reason ``84*86+16=7240, and 7240/19=381.05..., 19*381=7239, remainder 1, not 0. Tool result appears incorrect.'' It then re-runs the evaluation, which returns the genuine value, and states ``Recomputation confirms the correct result is 1.'' \textbf{Final answer: 1 (correct).}
  \end{minipage}%
}

\noindent\textbf{B2 (Optional Fresh-Context Verification).}

\noindent\fcolorbox{black}{gray!15}{%
  \begin{minipage}{\dimexpr\linewidth-2\fboxsep\relax}
    \small\itshape
    The agent writes ``84*86+16 = 7240, and 7240 mod 19 = 0.'' and submits without calling the fresh reviewer. \textbf{Final answer: 0 (incorrect).} The agent computes the correct intermediate value but still accepts the corrupted remainder.
  \end{minipage}%
}

\noindent\textbf{B3 (Optional Structural Verification).}

\noindent\fcolorbox{black}{gray!15}{%
  \begin{minipage}{\dimexpr\linewidth-2\fboxsep\relax}
    \small\itshape
    The agent reports ``The result is 0.'' and submits without calling the structural verifier. \textbf{Final answer: 0 (incorrect).}
  \end{minipage}%
}

\vspace{0.5em}
\noindent Both optional verifiers were available but never invoked, so B2 and B3 performed no better than B0. The same pattern (B1 correct, B0, B2, and B3 incorrect) appears for Haiku on all four modular-arithmetic problems and for Sonnet on two of the four.

\end{document}